\documentclass{PoS}

\PoS{PoS(BDMH2004)031}
\usepackage{epsfig}
\usepackage{natbib}

\newcommand{\HI}{{\protect\sc{HI}}}
\newcommand{\msun}{$M_\odot$}
\newcommand{\etal}{{et~al.}}
\newcommand{\mhi}{$M_{HI}$}
\newcommand{\kms}{km\,s$^{-1}$}

\title{What do loose groups tell us about galaxy formation?}

\ShortTitle{What do loose groups tell us about galaxy formation?}

\author{\speaker{D.J. Pisano}\thanks{Bolton Fellow, NSF MPS Distinguished
International Research Fellow}\\
        ATNF, P.O. Box 76, Epping, NSW 1710, Australia\\
        E-mail: \email{dj.pisano@csiro.au}}
\author{David Barnes\\ School of Physics, University of Melbourne, Victoria 
3010, Australia}
\author{Brad Gibson, Virginia Kilborn\\
Centre for Astrophysics \& Supercomputing, Swinburne University, Hawthorn, 
Victoria 3122, Australia}
\author{Lister Staveley-Smith\\ ATNF, P.O. Box 76, Epping, NSW 1710, Australia}
\author{Ken Freeman\\  RSAA, Mount Stromlo Observatory, Cotter Road, Weston, 
ACT 2611, Australia}

\abstract{We present the results of a Parkes Multibeam \HI\ survey of
six loose groups of galaxies analogous to the Local Group. This survey
was sensitive to \HI-rich objects in these groups to below
10$^7$\msun\ of \HI, and was designed to search for low mass, gas-rich
satellite galaxies and potential analogs to the high-velocity clouds
seen around the Milky Way.  This survey detected a total of 79
\HI-rich objects associated with the six groups, half of which were
new detections.  While the survey identified a small number of dwarf
galaxies, no star-free \HI\ clouds were discovered.  The \HI\ mass
function of the six groups appears to be roughly flat as is that of
the Local Group.  The cumulative velocity distribution function (CVDF)
of the \HI-rich halos in the six groups is identical to that of the
Local Group.  Both of these facts imply that these groups are true
analogs to the Local Group and that the Local Group is not unique in
its lack of low-mass dwarf galaxies as compared to the predictions of
cold dark matter models of galaxy formation.  This survey also
constrains the distance to and \HI\ masses of the compact
high-velocity clouds (CHVCs) around the Milky Way.  The lack of CHVC
analog detections implies that they are distributed within
$\lesssim$160 kpc of the Milky Way and have average \HI\ masses of
$\lesssim$4$\times$10$^5$\msun.  The spatial distribution of CHVCs is
consistent with the predictions of simulations for dark matter
halos.  Furthermore the CVDF of Local Group galaxies plus CHVCs
matches the predicted CVDF of cold dark matter simulations of galaxy
formation.  This provides circumstantial evidence that CHVCs may
be associated with low-mass dark matter halos.}

\FullConference{Baryons in Dark Matter Halos\\ 5-9 October 2004\\
		 Novigrad, Croatia}

\begin{document}
\section{Introduction}

Current models of hierarchical galaxy formation predict that galaxies
form via the accretion of smaller lumps of gas, stars, and dark matter
\citep[e.g][]{kau93}.  Simulations of this process assuming a cold
dark matter (CDM) universe uniformly predict the existence of large
numbers of low-mass dark matter halos surrounding massive galaxies
today \citep[e.g.][] {kly99,moo99}.  For the Local Group these authors
have shown that the simulations predict an order of magnitude more
dark  satellites than known luminous galaxies; a discrepancy described
as the ``missing satellite'' problem.  \citet{bli99} and \citet{bra99}
suggested that some or all of the ``high-velocity clouds''
\citep[HVCs; see][for a review]{wak97} may be associated with these
dark matter  halos and resolve this problem.  \citet{bli99} and
\citet{bra99} predicted that such HVCs would reside at distance of
$\sim$1 Mpc and have  \mhi$\sim$10$^7$\msun.  More recently
\citet*{deH02b} have predicted that only the compact HVCs (CHVCs) are
associated with dark matter halos, and that they are closer
(D$\sim$50-150 kpc) and less massive (\mhi$\sim$10$^{5-7}$\msun) than
the \citet{bli99} and \citet{bra99} predictions.   An
alternative possibility is that the Local Group is somehow unique in
its deficit of low-mass galaxies.  To test these hypotheses we have
been conducting an \HI\ survey of six spiral-rich loose groups
analogous to the Local Group using the Parkes Multibeam instrument to
search for low-mass, gas-rich dwarf  galaxies and analogs to the Local
Group HVCs.  These observations will serve as a benchmark for  the
\HI\ properties of galaxies within loose groups and a test of models
of galaxy formation.

The sample selection, parameters of the survey, and the initial
results from  the first half of the survey are described in
\citet{pis04b} and \citet{pis04c}, and will not be discussed here in great
detail.  This paper will  focus on the initial analysis of the entire
sample and the implications for the properties of loose groups,
galaxy formation, and the nature of HVCs.

\section{Group Properties}

Our sample of six loose groups was chosen from \citet{gar93} to only
contain spiral galaxies which were typically separated by $\sim$100
kpc and spread over an area of $\sim$1 Mpc$^2$.  All six groups are
located between 10-14 Mpc.  Our Parkes Multibeam survey on the
entirity of these groups was sensitive to objects with
\mhi$\gtrsim$10$^7$\msun.   We detected a total of 79 \HI-rich objects
in the six groups, half of which were previously unknown.  All
detections appear to have stellar counterparts.  While we have chosen
groups that are qualitatively similar to the Local Group, we wish to
determine if they are quantitatively analogous as well.  We will use
two measures to do this.  First, we will compare the shape of the
composite \HI\ mass function (\HI MF) of the six groups with that of
the Local Group.  Second, we will compare the cumulative velocity
distribution function (CVDF) of the Local Group and the loose groups.

\begin{figure}
\hfill\includegraphics[angle=-90,width=2.8in]{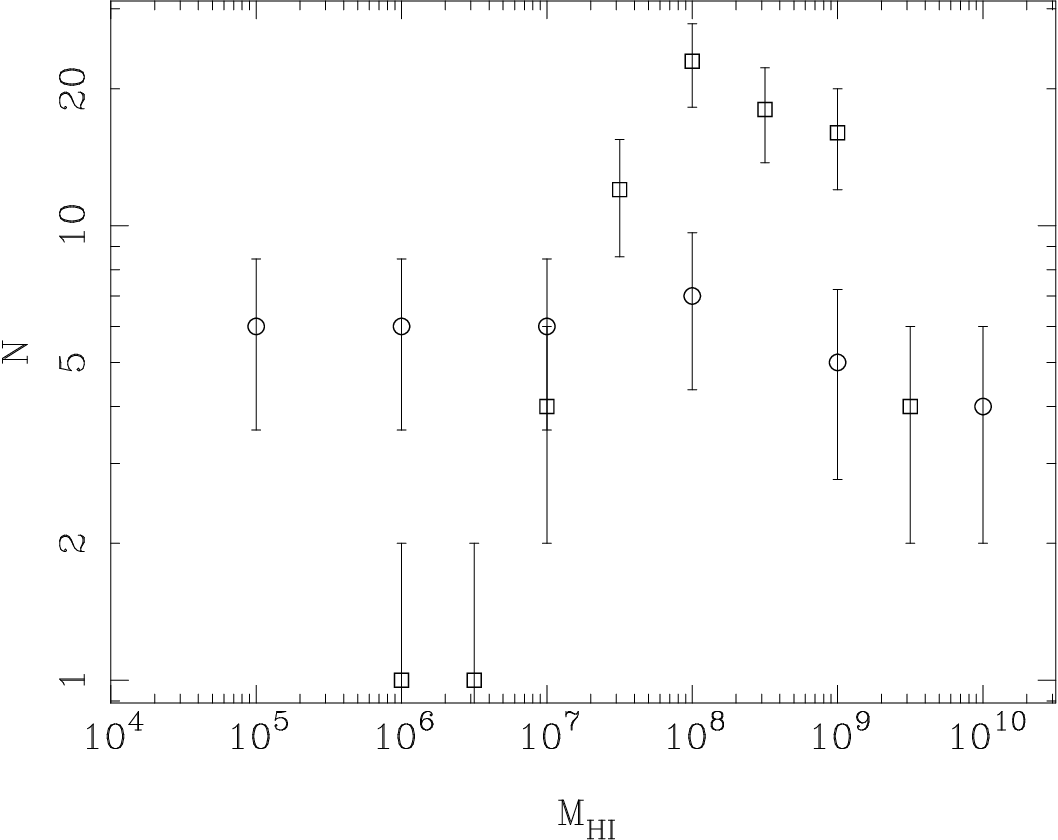}
\hfill\includegraphics[angle=-90,width=2.8in]{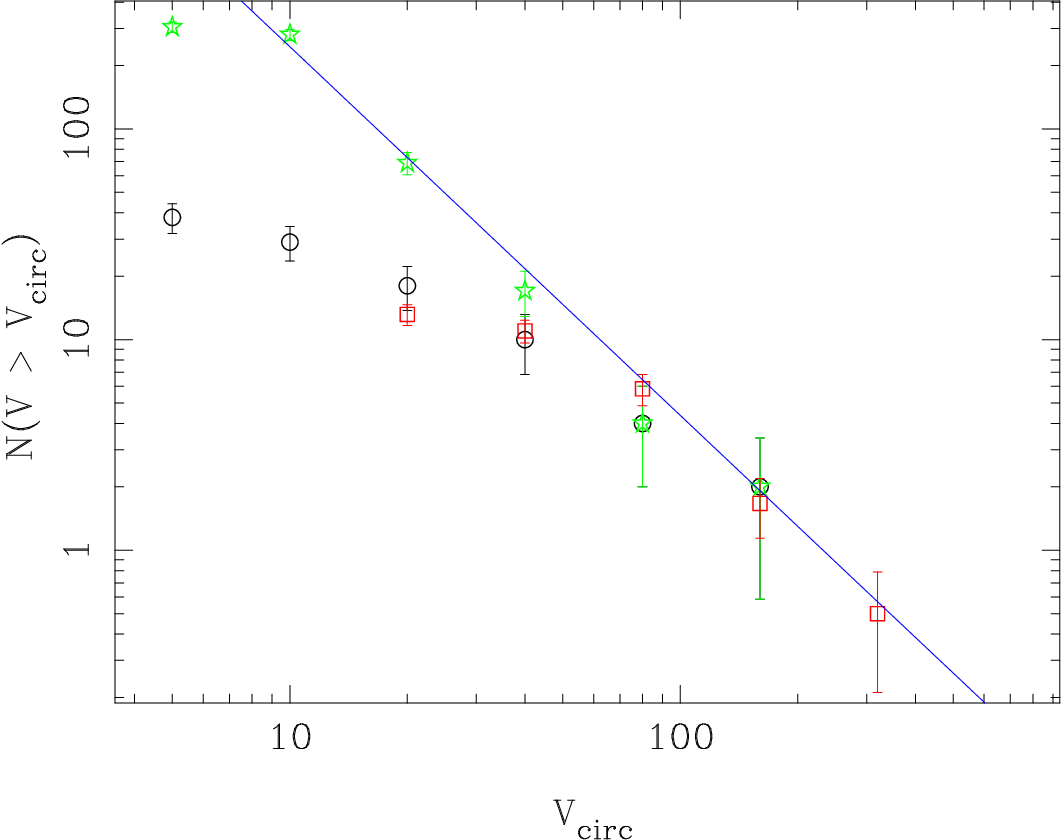}\hspace*{\fill}
\caption{(Left) The \HI\ mass function (\HI MF) of the Local Group
(black circles) and the sum of the six loose groups (red
squares).\label{fig:himf}}
\caption{(Right) The cumulative velocity distribution function for the
Local Group (black circles), the average of the six loose groups (red
squares), and the  Local Group including the CHVCs (green stars).  The
blue solid line represents the CDM model of Klypin roughly normalized
to the second data  point. \label{fig:cvdf}}
\end{figure}

Figure~\ref{fig:himf} shows the \HI MF  of the Local Group and the sum
of the six loose groups we observed.   The binned values simply
represent the total number of galaxies in each bin and not the volume
density of galaxies per dex.   Our group survey begins to be
incomplete for \mhi$<$10$^8$\msun, which explains the decline in
numbers across the last four bins for the loose groups.  Concentrating
instead on the four most massive bins, we can see that there is a
rough agreement in the slopes of the Local Group and loose group \HI
MFs.  This is still a very tentative result, which can be improved once we
correct the lower mass bins for incompleteness.   Generally, however,
we do see that the Local Group and other analogous loose groups have
similar \HI MFs.

One of the main goals of this project was to see if the Local Group
was unique in its relative lack of low-mass galaxies compared to the
predictions of cold dark matter models, such as those of \citet{kly99}
and \citet{moo99}.  In Figure~\ref{fig:cvdf}, we show the cumulative
velocity distribution function (CVDF) for the Local Group (black
circles)  and the average of the six loose groups (red squares) as
compared to the CDM simulation of \citet{kly99} marked with the blue
line.  The CVDF describes the number of dark matter halos with a
circular velocity greater than a given value; circular velocity is
used as a surrogate for mass.  It is immediately obvious that both the
Local Group and the loose  groups do not agree with the \citet{kly99}
simulations below $\sim$50 \kms, but agree almost perfectly with each
other.  Clearly whatever process is responsible for deficit of
low-mass luminous halos, be it alternative forms of dark matter
\citep*[e.g. warm dark matter][]{col00} or the suppression of baryonic
collapse \citep[e.g.][]{tul02} or something else is not unique to the
Local Group.  Furthermore, given the similarities in the \HI MF and
the CVDF between the Local Group and our sample of loose  groups, we
believe that our sample groups are truly analogous to the Local
Group.

\section{High Velocity Clouds and Dark Matter Halos}

Because our sample of loose groups appear to be true analogs to the
Local Group, we can constrain the distance and mass of the CHVCs
around the Milky Way.  In \citet{pis04b}, we determined that our
non-detection of analogs to CHVCs with the properties proposed by
\citet{bli99,bra99} down to our sensitivity limit implies that HVCs
must lie within D$\lesssim$160 kpc and have average \HI\ masses of
$\lesssim$4$\times$10$^5$\msun.  This is in concordance with a variety
of other observational and theoretical constraints \citep[see][for
further discussion]{pis04b}, and suggests that the CHVCs are more
closely associated with individual galaxies than with groups of
galaxies.

These distance and mass limits do not, however, rule out the
association of CHVCs with dark matter halos.  On the contrary, they
suggest a possible relation between CHVCs and  low-mass dark matter
halos.  Simulations show that dark matter halos are also distributed
within $\sim$100 kpc \citep*{kra04}.  In addition, if we use the HWHM of
the \HI\ linewidths of CHVCs from \citet*{deH02a} and \citet{put02},
and equate it to the circular velocity of a putative dark matter
halo, then we see on Figure~\ref{fig:cvdf} that CHVCs trace the CVDF
predicted by \citet{kly99} for CDM quite well.   The deviation from
CDM at small velocities is where the catalogs start to  become
incomplete.  This coincidence of the spatial and mass distribution of 
the CHVCs and low-mass CDM halos provides circumstantial evidence that
the CHVCs may be the missing dark matter satellites as originally suggested
by \citet{bli99}.


\begin{thebibliography}{99}

\bibitem[Blitz et al.(1999)]{bli99} Blitz, L., Spergel, D.N., Teuben, P.J., 
Hartmann, D., \& Burton, W.B., 1999, ApJ, 514, 818
\bibitem[Braun \& Burton(1999)]{bra99}Braun R., Burton W.B., 1999, A\&A, 
351, 437
\bibitem[Col\'in et al.(2000)Col\'in, Avila-Reese, \& Valenzuela]{col00} 
Col\'in, P., Avila-Reese, V., \& Valenzuela, O., 2000, ApJ, 542, 622
\bibitem[de Heij et al.(2002a)de Heij, Braun, \& Burton]{deH02a}de Heij V., 
Braun R., Burton W.B., 2002a, A\&A, 391, 67
\bibitem[de Heij et al.(2002b)de Heij, Braun, \& Burton]{deH02b}de Heij V., 
Braun R., Burton W.B., 2002b, A\&A, 392, 417
\bibitem[Garcia(1993)]{gar93}Garcia, A.M., 1993, A\&AS, 100, 47
\bibitem[Kauffmann, White, \& Guiderdoni(1993)]{kau93} 
Kauffmann, G., White, S.D.M., \& Guiderdoni, B., 1993, MNRAS, 264, 201
\bibitem[Klypin et al.(1999)]{kly99} Klypin, A., Kravtsov, A.V., Valenzuela, 
O., \& Prada, F., 1999, ApJ, 522, 82
\bibitem[Kravtov et al.(2004)Kravtsov, Gnedin, \& Klypin]{kra04}
Kravtsov, A.V., Gnedin, O.Y., Klypin, A.A., 2004, ApJ, 609, 402
\bibitem[Moore et al.(1999)]{moo99} Moore, B., Ghigna, S., Governato, F., 
Lake, G., Quinn, T., Stadel, J., \& Tozzi, P., 1999, ApJ, 524, L19
\bibitem[Pisano et al.(2004)]{pis04b}Pisano, D.J., Barnes, D.G., 
Gibson, B.K., Staveley-Smith, L., Freeman, K.C., \& Kilborn, V.A., 2004b, 
ApJ, 610, L17
\bibitem[Pisano(2004)]{pis04c}Pisano, D.J., 2004, PASA, 31, 392
\bibitem[Putman et al.(2002)]{put02}Putman, M.E., \etal, 2002, AJ, 123, 873
\bibitem[Tully et al.(2002)]{tul02}Tully, R.B., Somerville, R.S., 
Trentham, N., \& Verheijen, M.A.W., 2002, ApJ, 569, 573
\bibitem[Wakker \& van Woerden(1997)]{wak97} Wakker, B.P., \& van Woerden, H., 
1997, ARA\&A, 35, 217

\end{thebibliography}
\end{document}